\documentclass[twocolumn]{aa}
\usepackage{graphicx}
\usepackage{txfonts}
\usepackage{natbib}
\begin{document}
   \title{Study of the cyclotron feature in \object{MXB 0656-072}}

   \author{V.A. McBride\inst{1}
          \and J. Wilms\inst{2}
	  \and M.J. Coe\inst{1}
	  \and I. Kreykenbohm\inst{3}$^,$\inst{4}
	  \and R.E. Rothschild\inst{5}
	  \and W. Coburn\inst{6}
	  \and J.L. Galache \inst{1}
	  \and P.~Kretschmar \inst{7}
	  \and W.R.T. Edge \inst{1}
	  \and R. Staubert \inst{3}
          }

   \offprints{V.A.McBride \\\email{vanessa@astro.soton.ac.uk}}

   \institute{School of Physics \& Astronomy, University of Southampton,
     Highfield, SO17 1BJ, UK
               \and Department of Physics, University of Warwick,
     Coventry CV4 7AL, UK
               \and Institut f\"ur Astronomie und Astrophysik --
               Astronomie, Sand 1, 72076 T\"ubingen, Germany
	       \and INTEGRAL Science Data Centre, 16 ch. d'\'Ecogia,
               1290 Versoix, Switzerland
	       \and Center for Astrophysics and Space Sciences,
               University of California at San Diego, La Jolla, CA
               92093-0424, USA
	       \and Space Sciences Laboratory, University of California
             at Berkeley, Berkeley, CA 94702-7450, UK
               \and European Space Astronomy Centre (ESAC),
             European Space Agency, P.O. Box 50727, 28080, Madrid, Spain
             }

   \date{Received September 23, 2005; accepted January 26, 2006}

   \abstract{
    We have monitored a Type II outburst of the Be/X-ray binary
   \object{MXB 0656$-$072} in a series of pointed \emph{RXTE}
   observations during October through December 2003.  The source spectrum
   shows a cyclotron resonance scattering feature at $32.8^{+0.5}_{-0.4}$\,keV,
   corresponding to a magnetic field strength of $3.67^{+0.06}_{-0.04}\times
   10^{12}$\,G and is stable through the outburst and over the pulsar
   spin phase.  The pulsar, with an average pulse period of
    $160.4\pm0.4$\,s, shows a spin-up of 0.45\,s over the duration of
    the outburst.   From optical data, the source distance is
    estimated to be $3.9\pm0.1$\,kpc and this is used to estimate the
    X-ray luminosity and a theoretical prediction of the pulsar
    spin-up during the outburst.

   \keywords{X-rays: stars 
     -- stars: magnetic fields
     --stars: pulsars: individual: MXB 0656$-$072
               
               }
   }

   \maketitle

\section{Introduction}

\object{MXB 0656$-$072} was first classified as a transient source by
\citet{cs75} when detected at 80\,mCrab on 1975 September 20 by
\emph{SAS-3}.  Subsequently \emph{Ariel V} observed the source at 50
and 70\,mCrab 
on 1976 March 19 and 27
respectively \citep{k76}.  Kaluzienski's comments that \object{MXB 0656$-$072}
may more closely resemble a long term variable source as opposed to a
transient possibly led to it being classified as a Low Mass X-ray Binary in the
catalogue of \citet{lp01}.

The source, which has been dormant and unobserved since 1975/1976,
made a reappearance in the X-ray sky in a large, extended outburst
during 2003 October.  During this outburst,which lasted over 2 months
and reached an X-ray luminosity of 200\,mCrab, it was identified as a
pulsating X-ray binary with a pulse period of 160.7\,s \citep{mr03} and
with an optical counterpart of spectral type O9.7Ve \citep{pm03},
re-categorising it as a High Mass X-ray Binary.  Time-resolved B and R
photometry \citep{BartoliniCasaresGuarnieri2005} of the optical
counterpart during this outburst revealed no optical periodicity at the X-ray
pulse period.  Orbital parameters of the system remain undetermined.

Preliminary analysis of the phase averaged spectrum by \citet{hc03} showed that the continuum could be fit with a power law with an
exponential cutoff.  A cyclotron resonant scattering feature with a
centroid energy of 36$\pm$1\,keV was discovered by \citet{hc03}.

The strong magnetic fields found near the polar caps of neutron stars
quantise the ambient electron energies into Landau levels.  Photons
at these resonant electron energies are scattered, creating cyclotron
resonant scattering features (CRSFs) in the observed spectrum.  The energy of the fundamental CRSF is approximated by:
\begin{equation}
E_{\rm C}\simeq 11.6\,{\rm keV}\times \frac{1}{1+z}\times \frac{B}{10^{12}\rm{G}}
\end{equation}
where $z$ is the gravitational redshift and $B$ the magnetic field strength.  Thus measuring the energy of cyclotron features in the spectra of accreting X-ray binaries gives us a direct measurement of the magnetic field of the neutron star.\\

\section{Data Reduction}

\begin{figure*}
\includegraphics[width=12cm]{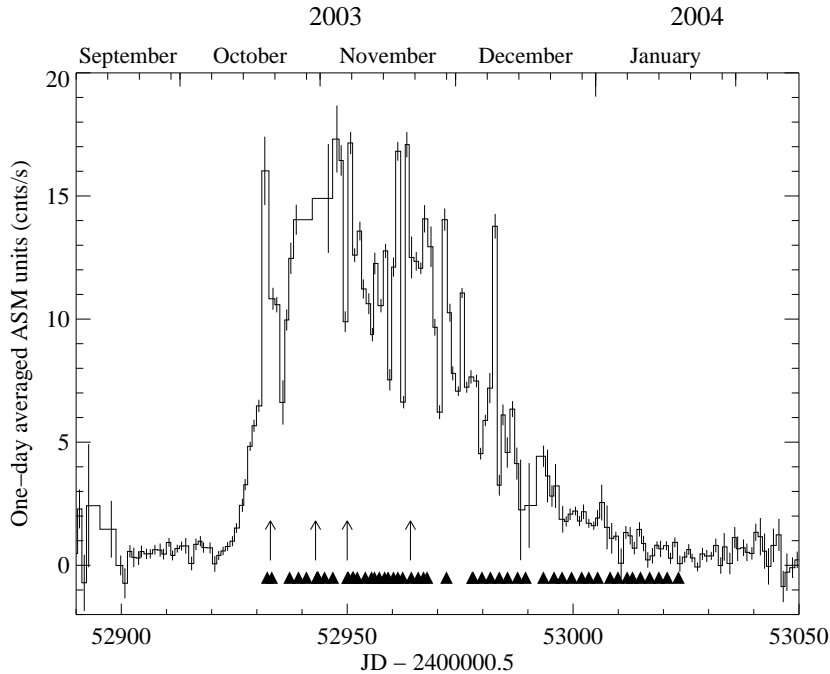}
\caption{The ASM lightcurve of \object{MXB 0656$-$072} during the 2003
  outburst with bin sizes of 1 day. Triangles indicate the dates of pointed \emph{RXTE}
  observations. The arrows indicate the dates at which spectra were
  compared with each other in order to observe any spectral evolution
  over the outburst.  This is further discussed in Sect.~3.2.}
              \label{FigAsm}
\end{figure*}

\object{MXB 0656$-$072} was observed with \emph{RXTE} during 2003 October through 2004 January.  Observations coincided with a Type II \citep{sw86} X-ray
outburst and data were accumulated during both the rise and decline of
the outburst.  In total, the \emph{PCA} livetime for this set of observations
amounts to 95\,ks. Fig.~\ref{FigAsm} shows the \object{MXB 0656$-$072}
lightcurve from the All Sky Monitor (\emph{ASM}) aboard \emph{RXTE} during the
outburst. Triangles indicate the dates of pointed \emph{RXTE} observations.

While \emph{Standard2f} mode \emph{PCA} data (with 16\,s time resolution and 128 channel
energy resolution) were employed to generate phase averaged spectra,
the phase resolved spectra and timing results were obtained from observation-specific binned-mode data with a
temporal resolution of 250\,ms and an energy resolution of 128
channels.

Analysis was performed with \emph{HEASOFT} version 5.3.1 and spectral
fitting with \emph{XSPEC} version 11.3.1t. Phase resolved spectra were
created using a modified version of the \emph{FTOOL fasebin} developed
at the University of T\"ubingen \citep{kreykenbohm2004}.

For spectral fitting, \emph{PCA} data \citep{JahodaMarkwardtRadeva2005} in the energy range 3--22\,keV were used,
to which we added systematic errors of 0.5$\%$.  Below and above this
range the systematic errors required are larger to account for the
uncertainties in the \emph{PCA} response matrix, so we ignored data
outside the 3--22\,keV range.

For all spectra we used the E\_$8\mu s$\_256\_DX1F mode \emph{HEXTE}
data \citep{RothschildBlancoGruber1998}, which have
a temporal resolution of 8\,$\mu$s and covers the range 0--250\,keV
with 256 spectral channels.  Electronic thresholds limit the
\emph{HEXTE} energy range to 15--250\,keV.  In all cases, the
signal-to-noise ratio was increased by adding together the data from
both clusters and generating a new response file with weights 1:0.75
to account for the loss of signal from one of the detectors in the
second \emph{HEXTE} cluster. A binning factor of 3 was applied to data
in the 70--200\,keV range to improve statistical significance of model
fitting while still maintaining a reasonable energy resolution.  Data
above 200\,keV were not used.

\section{Phase Averaged Analysis}

For the phase averaged spectral analysis, the spectra from all
observations through the peak of the outburst (from MJD 52932 to MJD 52964)  were added together in an effort to maximise the signal-to-noise ratio. 

\subsection{Spectral Model}
In accreting X-ray pulsars the highly ionised accretion stream is
channelled onto one or both of the magnetic poles of the neutron star by its strong
($B\sim 10^{12}$\,G) magnetic field.  Depending on the accretion rate
and the strength of the magnetic field, an accretion shock may form.  The continuum spectral shape is
dominated by Comptonisation of photons by the accretion stream at the
magnetic poles.  As a comprehensive theoretical spectral model
is unavailable, the spectrum is generally modelled as a exponentially
cutoff power law, but see  \citet{t86} for the choice of a Fermi-Dirac
Cut Off (FDCO) model or \citet{m95} for the NPEX model. In addition to this continuum we may observe photoelectric absorption by the accretion stream or stellar wind, an Fe
K$\alpha$ fluorescence line and one or more cyclotron resonance
scattering features.

In our analysis we describe the continuum using a power law with photon
index $\Gamma$, with a high energy cutoff at $E_{\rm cut}$. 
\begin{equation}
I_{\rm cont}(E)=KE^{-\Gamma}\times\left\{\begin{array}{ll}
1 & (E\leq E_{\rm cut})\\
e^{-(E-E_{\rm{cut}})/E_{\rm{fold}}} & (E > E_{\rm cut})
\end{array}\right.
\end{equation}
 The energy cutoff is smoothed by including a Gaussian absorption
 line (\emph{gsmooth} in \emph{XSPEC}) with an energy dependent width, at the continuum cutoff
 energy.  Without a smoothing function the energy cutoff forms a
 discontinuity in the model and this results in residuals which can easily be
 mistaken as arising from a cyclotron feature
 \citep{KretschmarPanKendziorra1997,c01}.  Although we attempted to
 use the FDCO and NPEX models to describe the continuum, we found that
 the best-fit continuum was indeed a power law with a smoothed
 high-energy cutoff.
  
The spectral shape is insensitive to the H I column density and in order
to include some estimate of the column density a value
of $7.28 \times 10^{21}$\, cm$^{-2}$ was estimated over a cone of
radius $1^{\circ}$ in the direction of the source
\citep{DickeyLockman1990}.  The column density, $N_{\rm{H}}$, was
fixed at this value for subsequent spectral fitting. 

Modifying the continuum is a CRSF modelled
with the CYCLABS model, which has the analytical form shown below:
\begin{equation}
CYCLABS(E)= D_{\rm c}\frac{(W_{\rm c}E/E_{\rm c})^2}{(E-E_{\rm c})^2+W_{\rm c}^2}
\end{equation}
\noindent
where $E_{\rm c}$, $D_{\rm c}$ and $W_{\rm c}$ are the cyclotron energy, depth and width
respectively \citep{mo90,mm90}.  Using other continuum models (FDCO
and NPEX) did not alter the presence of the cyclotron feature.

We model a strong Fe K$\alpha$ fluorescence feature with a
Gaussian emission line and determine the equivalent width to be
3.66\,keV.  In addition we include a blackbody component to account
for the excess flux at the soft end of the spectrum. The blackbody
component has a temperature of $0.86^{+0.08}_{-0.14}$\,keV and can be
attributed to emission from an accreting polar cap, in a manner similar to
\citet{CoburnHeindlGruber2001} in the case of \object{X Per}. 

\begin{figure*}
\includegraphics[width=12cm]{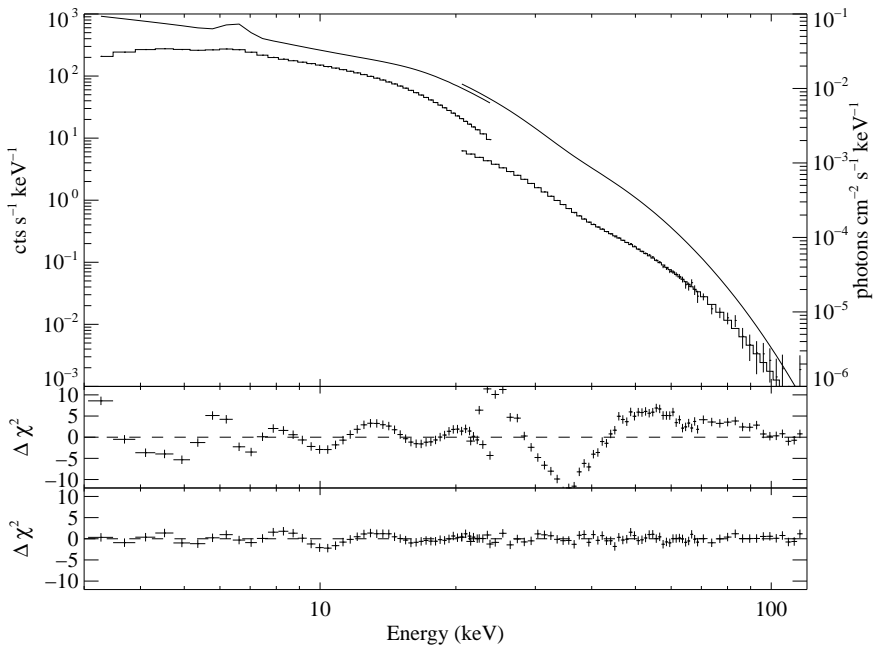}
\caption{The spectrum of MXB 0656$-$072 from 3--120 keV.  The crosses
  show the data, the smooth curve shows the unfolded spectrum and the
  histogram the model fit.  In the second panel the residuals, plotted
  as $\Delta\chi^2$, are shown for the case in which no CRSF is
  included in the model.  In the lower panel, these residuals are
  shown again -- this time including a CRSF in the model.}
              \label{FigPas}
\end{figure*}

The fit to the phase averaged spectrum results in
$\chi^2_{\rm{red}}=1.1$ with the spectral parameters given in Table
~\ref{TabSpecp}.  The spectrum is shown in Fig.~\ref{FigPas}.

 \begin{table}
      \caption[]{Spectral parameters (as defined in the text) for the phase averaged spectrum.
      The model consists of power law with a smoothed high energy
      cutoff, photoelectric absorption, a CYCLABS cyclotron feature, a
      Gaussian Fe line and a blackbody component.  The flux in the Fe
      emission line is represented by $a_{\rm Fe}$, in $\textrm{photons}\,\textrm{cm}^{-2}\textrm{s}^{-1}$.  Uncertainties are at 90\% confidence intervals.}
         \label{TabSpecp}
	 \centering
         \begin{tabular}{ll}
            \hline\hline
            \noalign{\smallskip}
            \textrm{Parameter}      &  \textrm{Value}  \\
            \noalign{\smallskip}
            \hline
	    \noalign{\smallskip}
	    $N_{\rm{H}}$ & $7.28\times 10^{21}$ cm$^{-2}$\\
            \noalign{\smallskip}
	    $\Gamma$ & $0.89^{+0.07}_{-0.05}$\\
	    \noalign{\smallskip}
	    $E_{\mathrm{cut}}$ & $15.5^{+0.2}_{-0.2}$ keV\\
	    \noalign{\smallskip}
	    $E_\mathrm{fold}$ & $11.8^{+0.2}_{-0.3}$ keV\\
	    \noalign{\smallskip}
	    $T$ & $0.86^{+0.08}_{-0.14}$ keV\\
	    \noalign{\smallskip}
	    $E_\mathrm{c}$ & $32.8^{+0.5}_{-0.4}$ keV\\
	    \noalign{\smallskip}
	    $W_\mathrm{c}$ & $11.8^{+1.0}_{-1.1}$ keV\\
	    \noalign{\smallskip}
	    $ D_\mathrm{c}$ & $0.38^{+0.02}_{-0.02}$\\
	    \noalign{\smallskip}
	    $E_\mathrm{Fe}$ & $6.48^{+0.03}_{-0.03}$ keV\\
	    \noalign{\smallskip}
	    $\sigma_\mathrm{Fe}$ & $0.36^{+0.06}_{-0.06}$ keV\\
	    \noalign{\smallskip}
	    $a_\mathrm{Fe}$ & $0.0195^{+0.001}_{-0.001}$\\
            \noalign{\smallskip}
            \hline
         \end{tabular}
 
\end{table}

The cyclotron feature at $32.8^{+0.5}_{-0.4}$\,keV is clearly visible
in the residuals.  A harmonic with width $16^{+20}_{-14}$\,keV and
depth $0.19^{+0.26}_{-0.16}$ may be present at twice the fundamental
energy, but detection is marginal, as can be interpreted from the
uncertainties in the line parameters.
Assuming that the cyclotron line at 32.8\,keV is the fundamental ($n=1$) and a
gravitational redshift of $z = 0.3$ for a typical neutron star mass of $1.4\,M_{\odot}$ and radius of 10\,km, we can estimate the magnetic field from Eq.1 to be $3.67^{+0.06}_{-0.04}\times 10 ^{12}$\,G.

Note the systematic feature in the residuals at $\sim$10\,keV.  This
feature is evident in the spectra of a number of accreting X-ray
binaries (\object{GS 1843+00}, \object{Her X-1}) \citep{c01}.  It has been
 observed in spectra accumulated by \emph{Ginga} \citep{m95} and
\emph{Beppo}SAX \citep{sd98} as well, and should be accounted for
in future continuum models of accreting X-ray pulsars.  A more
comprehensive review of this feature can be found in \citet{c01}.

\subsection{Spectral Evolution over Outburst}

In order to determine whether there was any significant change in the
spectral shape or the cyclotron feature during the outburst we
compared the spectral parameters at various dates through the
outburst.  These dates are indicated by the
arrows on Fig.~\ref{FigAsm}  and further information is shown in Table ~\ref{TabSev}.

 \begin{table}
      \caption[]{Dates and \emph{PCA} livetimes of observations used
      to determine the spectral evolution over the outburst.}
         \label{TabSev}
	 \centering
         \begin{tabular}{l l l}
            \hline\hline
            \noalign{\smallskip}
            Date & MJD &  Livetime (ks)  \\
            \noalign{\smallskip}
            \hline
	    2003-10-20&52933&11\\
	    2003-10-30&52943&8\\
	    2003-11-06&52950&15\\
	    2003-11-(16-20)&5296(0-4)&7\\            
            \noalign{\smallskip}
            \hline
         \end{tabular}
 
\end{table}

\begin{figure}
\centering
\includegraphics[angle=0,width=8cm]{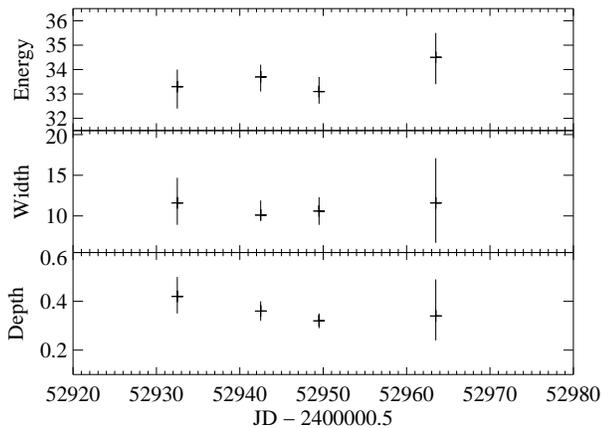}
\caption{Cyclotron line parameters throughout the 2003 October
  outburst. The energy and width are in keV.}
              \label{FigSpecev}
\end{figure}

Note that the fourth spectrum is obtained by adding observations over
a five day range in order to have signal-to-noise ratios similar to
those of the first 3 datasets.  We find that, on average, the spectral
fits have $\chi_{\rm red} ^2 \sim1.2$, except for 2003 November
16--20 which has $\chi_{\rm red}^2 = 1.7$.  The power law index stays
between 0.9 and 1.1 during this time and there is very little change
in $E_{\rm cut}$ and $E_{\rm fold}$, which have values close to those
given in Table ~\ref{TabSpecp}.  The cyclotron parameters over this
time interval are shown in Fig.~\ref{FigSpecev}.  Noting that the error bars
represent 90\% uncertainty levels it is clear that the cyclotron
feature is stable over the outburst.

\section{Phase Resolved Analysis}

If the rotation and magnetic axes of the neutron star are not aligned, the observer sees X-ray emission modulated by the pulsar spin period.  Due to the fact that cyclotron features are strongly dependent on viewing angle, phase resolved spectroscopy is a tool well-suited to studying cyclotron features.

\begin{figure}
\centering
\includegraphics[angle=0,width=9cm]{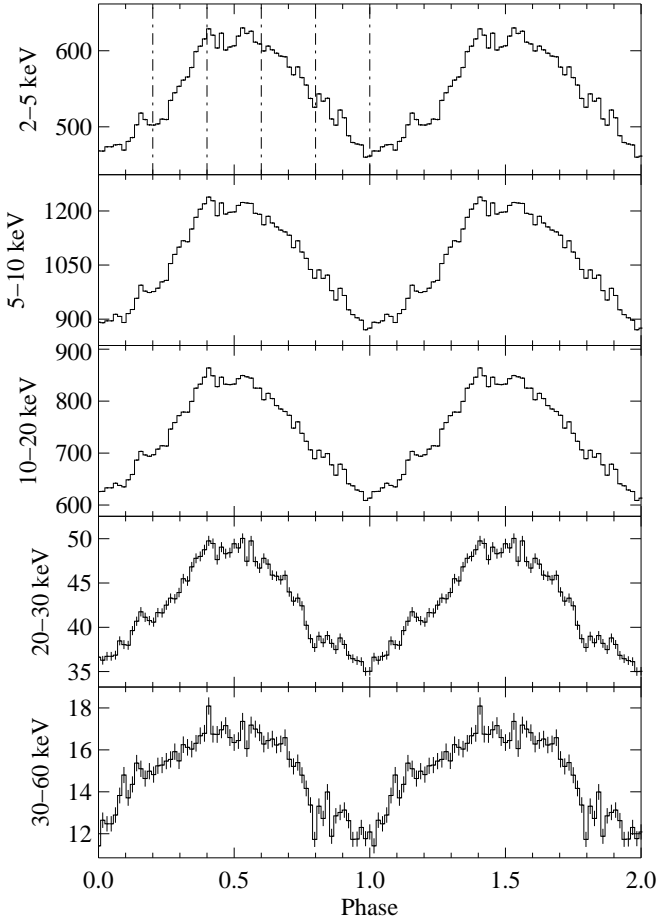}
\caption[]{Pulse profile shown in  counts s$^{-1}$.  Profiles are plotted twice for clarity.}
\label{FigPulsep}
\end{figure}

Pulse profiles and phase resolved spectra were generated by performing
a barycentric correction of the bin times for the \emph{PCA} data and
the photon arrival times for the \emph{HEXTE} data for two 15\,ks observations at the peak of the outburst.  Orbital
modulation was not accounted for as the orbital ephemeris of the
system is still unknown.  The pulse period was determined by running a
Lomb-Scargle analysis \citep{PressRybicki1989,Lomb1976,Scargle1982} on the barycentre-corrected lightcurves.  The
lightcurves were then folded at the pulse period of 160.4\,s to produce the pulse profiles shown in Fig.~\ref{FigPulsep}.

\subsection{Pulse profiles}
The pulse profiles are single-peaked and sinusoidal through all energy bands
 and show the pulsed fraction ($(F_{\rm max}-F_{\rm min})/F_{\rm max}$)
increasing steadily from 26\% in the 2--5\,keV band up to 40\% in the
30--60\,keV band.  The pulse shape is consistent with what one would
 expect from viewing just a single magnetic pole radiating isotropically as a
 blackbody, including the effects of gravitational bending around the
 neutron star \citep{beloborodov2002}. The almost perfectly sinusoidal
 pulse shapes may indicate that there is no absorption along the accretion
 column, as it is believed that strong absorption along the accretion
 column contributes to the complexity of the pulse profile at low energies
 \citep{Kreykenbohm2002,NagaseHayakawaMakino1983}.  The pulse profile
 shape and amplitude support the conclusion reached by
 \citet{BulikGondek-RosinskaSantangelo2003}:  that the rotation and magnetic axes of accreting pulsars are almost aligned.

\subsection{Spectral evolution over Pulse}
\begin{figure}
\centering
\includegraphics[angle=0,width=8cm]{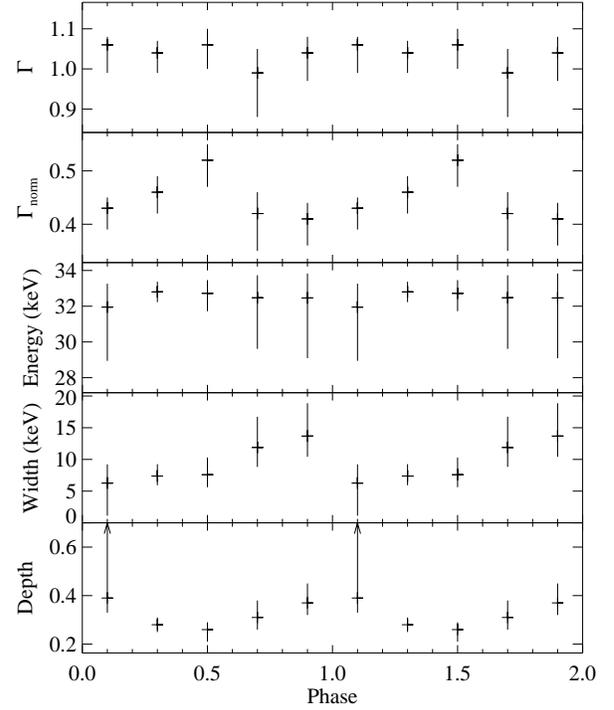}
\caption{From top to bottom the power law index, the power law
  normalisation, the cyclotron energy, cyclotron line width and
  cyclotron line depth are plotted as a function of phase.  Error
  bars indicate 90\% uncertainty levels.}
              \label{FigPhaseParam}
\end{figure}

The data were split into 5 phase bins for analysis, as shown in the
topmost panel of Fig.~\ref{FigPulsep}.  This gives us good statistics
at a temporal resolution sufficient to describe this particular pulse
shape.  Evolution of the spectral
parameters with phase is shown in Fig.~\ref{FigPhaseParam}.

For the phase resolved spectra, we used the same model as in
Sect.~3.1, however the addition of a blackbody component did not
significantly improve the fit, so was not included. Furthermore, a
Gaussian absorption component at $\sim$10\,keV was required to take
account of the systematic feature at this wavelength, mentioned in Sect.~3.1.  

The cyclotron feature at energy around 32\,keV is present through all
pulse phases. There are, however, variations in the cyclotron line width.  In general, the line is at
its narrowest ($6^{+3}_{-5}$\,keV) in the peak rise and reaches its broadest
($14^{+5}_{-3}$\,keV) during the pulse decline (See Fig.~\ref{FigPhaseParam}.)  The cyclotron line depth
also shows some variation through phase, in a manner inverse to the
continuum power law normalisation coefficient.  However, the cyclotron feature in this source shows little variability with phase in comparison with some other sources such as \object{Vela X-1} \citep{Kreykenbohm2002} and \object{GX 301-2} \citep{KreykenbohmWilmsCoburn2004}, which show cyclotron line energy variations of a couple keV throughout the phase. 

\section{Discussion}
\subsection{Source Distance}

As the optical counterpart to \object{MXB 0656$-$072} has been
identified, we can use the optical magnitude and colour to estimate
the distance to the source.

The optical counterpart has been identified as an O9.7Ve spectral
type star, with an observed $B-V$ colour in the range 0.86 to 1.02
\citep{pm03} and an observed $V$-magnitude in the range 12.05\,mag to
12.38\,mag. For this estimate we shall assume that the variability in the
$V$-magnitude and the $B-V$ colour is due to the circumstellar disk
contributing the the $V$-magnitude.  Hence we will use the faintest
available $V$-magnitude value, assuming that this is when the disk is
smallest, along with the corresponding bluest $B-V$ colour.  We have
estimated errors typical for photometry of a 12th magnitude star on a
1--2\,m telescope and use $V=12.38 \pm 0.02$\,mag and $B-V=0.86 \pm 0.03$.

Interstellar reddening is given by:
\begin{equation}
E(B-V)=(B-V)-(B-V)_0
\end{equation}
where $(B-V)_0$ is the intrinsic colour.  Thus for an O9.7Ve star we
expect $(B-V)_0=-0.28$ \citep{Wegner1994} and an absolute magnitude of $M_V=-4.2$\,mag \citep{z90}.

This puts the magnitude of reddening in the $V$-band at $A=3.7 \pm
0.1$\,mag. Knowing the apparent magnitude, the absolute magnitude and
the reddening magnitude allows us to estimate the distance using:
\begin{equation}
D  = 10^{1+(m-M-A)/5} =  3.9\pm 0.1\,\rm{kpc}
\end{equation}

It is most likely that the optical colour and magnitude used in this
calculation were observed at time when the contribution from the Be
star disk was not negligible.  Hence the above distance estimate will
be an underestimate.

\subsection{Spin-up trends}

Accretion of material onto a pulsar is expected to transfer a
significant amount of angular momentum to the neutron star.  Thus an
accreting X-ray pulsar undergoing an outburst is in a state where the
mass (and thus momentum) transfer rate is high and one expects to observe a
slight decrease in the pulse period as it is spun-up. 

\begin{figure}
\centering
\includegraphics[angle=0,width=8.5cm]{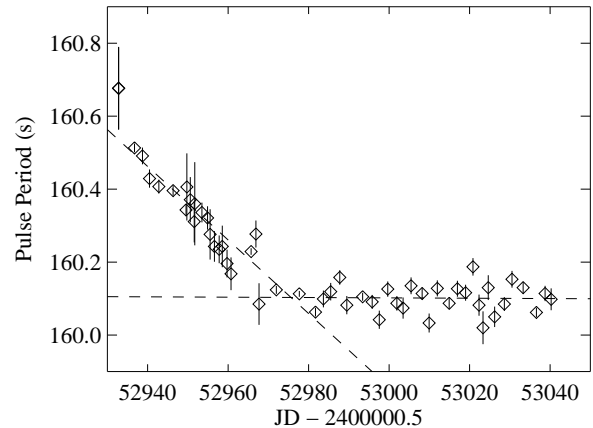}
\caption{Pulse period as a function of time through the outburst, with a linear fit to the data.}\label{FigPeriodTime}
\end{figure}

We have analysed all pointed \emph{PCA} data in the 2--20\,keV energy range
and determined pulse periods as a function of time.  Initially, to
derive the pulse period, we used a Lomb-Scargle analysis
\citep{PressRybicki1989}.  Since the resulting values had a
rather high uncertainty, the method of phase connection
\citep{DeeterPravdoBoynton1981,MunoChakrabartyGalloway2002} was
applied in a second step:  average pulse profiles (with an initial
period of $P=160.4$\,s) for data sets of individual \emph{RXTE} orbits
(of typical length 1 to 3\,ks) were produced and phase connected by an
integer number of pulse cycles. Using an improved solution for the
pulse ephemeris, including $\dot P$, also the larger gaps between data
sets could be uniquely bridged. The result is shown in
Fig.~\ref{FigPeriodTime}.  The measurements are consistent with an
initial period of 160.58\,s at MJD 52930 and an average spin-up until
MJD 52971 of $\dot P = -0.0101 \pm 0.0003\, \textrm{s day}^{-1}$.

After MJD 52971 the accretion rate drops to a level low enough to halt
spin-up of the neutron star.  This is evidenced by a flattening of the
slope in the pulse period vs time plot shown in
Fig.~\ref{FigPeriodTime}.  The data after MJD 52971 are consistent
with a constant period.  This levelling off of the pulse period to a constant value after MJD
52971 indicates that the pulse period changes are an effect of
accretion torques rather than modulation by the orbital period of the neuton star.

From the average X-ray flux in the 2--10\,keV band ($4.35\times 10^{-9}$\,erg
cm$^{-2}$ s$^{-1}$) we can calculate the X-ray luminosity if the
distance to the source is known.  Using our estimate of $3.9\pm
0.1$\,kpc for the source distance from Sect.\,5.1, the X-ray
luminosity of the source is $6.6 \pm 0.4 \times 10^{36}$\,erg
  s$^{-1}$.  Then a straightforward application of the accretion
    torque model by \citet{GhoshLamb1979} (Eq.\,(15) with standard
    neutron star parameters and our own values of $B = 3.67 \times
    10^{12}$\,G and $L_{\mathrm x}=6.6\times
10^{36}$\,erg s$^{-1}$) yields a spin-up rate of $\dot P=-0.0034\,\textrm{s
  day}^{-1}$.  This is about a factor
of three smaller than the observed value. We do not consider this a
problem, as accreting binary pulsars very generally show a
broad range of behaviour with respect to the development of their spin
periods with time (for reviews see \citet{Nagase1989} and
\citet{BildstenChakrabartyChiu1997}). This includes episodes of
spin-up and spin-down, depending on the exact conditions of accretion
at the time. For a transient in outburst such as MXB 0656-072 one
would expect to observe a strong spin-up, since under the greatly
increased mass accretion rate, as evidenced by the large increase in
X-ray luminosity, the system will tend to be far from any equilibrium
condition.

\section{Summary}
From analysis of the pointed \emph{RXTE} observations of \object{MXB
  0656$-$072} and limited optical data available we can deduce the
following:
\begin{itemize}
\item The X-ray spectrum continuum is best fit by a power law
  model with an exponential cutoff, where a Gaussian function with an
  energy dependent width is used to smooth discontinuities at the
  cutoff energy . 
\item A cyclotron resonance scattering feature is present at an energy
  of $32.8^{+0.4}_{-0.5}$\,keV and the energy is independent of the
  continuum model used to describe the spectrum.  The CRSF is present
  at this energy in all observations throughout the outburst.  The
  CRSF is present throughout the entire neutron star spin phase, but
  the width and depth of the feature show variation through the spin phase. 
\item Although there is no convincing evidence for higher harmonic cyclotron lines in this data, the possibility of a higher harmonic cannot be ruled out.
\item The magnetic field inferred  from the cyclotron energy (by
  assuming that the detected line is the fundamental CRSF) is
  $3.67^{+0.04}_{-0.06}\times 10^{12}$\,G.
\item On the basis of optical photometric data, the distance to
  \object{MXB 0656$-$072} is estimated to be $3.9\pm0.1$\,kpc.
\item Over the time considered, we have measured a strong spin-up
  (about 0.45\,s in 30\,days), which appears quite reasonable under the
  conditions observed in this high mass X-ray binary in outburst.
\end{itemize}

\begin{acknowledgements}
VAM would like to acknowledge the NRF(S.Africa), the British Council
and Southampton University. IK acknowledges DLR grants 50 OG 9601 and 50 OG 0501.
\end{acknowledgements}

\bibliographystyle{aa}
\bibliography{b4239}
\end{document}